\documentstyle{article}
\def\ve{\varepsilon}

\def\F{{\cal F}}

\def\der{\partial}
\def\bq{\begin{equation}}
\def\eq{\end{equation}}
\def\brr{\begin{eqnarray}}
\def\err{\end{eqnarray}}
\def\ba{\left(\begin{array}}
\def\ea{\end{array}\right)}
\def\dslash{\hbox{\ooalign{$\displaystyle\partial$\cr$/$}}}

\newcommand{\dr}{\raise.3ex\hbox{$\stackrel{\leftarrow}{\partial }$}{}}
\newcommand{\dl}{\raise.3ex\hbox{$\stackrel{\rightarrow}{\partial}$}{}}

\newcommand{\ft}[2]{{\textstyle\frac{#1}{#2}}}

\renewcommand{\d}{\delta}

\newcommand{\g}{\gamma}

\newcommand{\e}{\epsilon}

\renewcommand{\l}{\lambda}

\newcommand{\m}{\mu}

\newcommand{\n}{\nu}

\newcommand{\s}{\sigma}

\def\1ad{\mbox{\normalsize $^1$}}
\def\2ad{\mbox{\normalsize $^2$}}
\def\3ad{\mbox{\normalsize $^3$}}
\def\4ad{\mbox{\normalsize $^4$}}
\def\5ad{\mbox{\normalsize $^5$}}
\def\6ad{\mbox{\normalsize $^6$}}
\def\7ad{\mbox{\normalsize $^7$}}
\def\8ad{\mbox{\normalsize $^8$}}
\def\makefront{\vspace*{1cm}\begin{center}
\def\newtitleline{\\ \vskip 5pt}
{\Large\bf\titleline}\\
\vskip 1truecm
{\large\bf\authors}\\
\vskip 5truemm
\addresses
\end{center}
\vskip 1truecm
{\bf Abstract:}
\abstracttext
\vskip 1truecm}
\begin{document}
%%%%%%%%%%%%%%%%%%%%%%%%%%%%%%%%%%%%%%%%%%%%%%%%%%%%%%%%%%%%%%%%%%%%%%%%%%%%%%
\begin{titlepage}
\begin{flushright} KUL-TF-98/06\\ THU-98/05\\ HUB-EP-98/06 \\
ITP-SB-98-12 \\ DFTT-3/98 \\[3mm]
{\tt hep-th/9801200}
\end{flushright}
\vspace{16mm}
\begin{center}
{\LARGE\bf Vector-tensor multiplets\footnote{Based on a talk given by P.
Claus at the
Workshop on Quantum Aspects of Gauge Theories, Supersymmetry and
Unification, Neuchatel, September 18--23, 1997; to be published in the
proceedings.}}
\vfill
{\large P. Claus$^1$,
        B. de Wit$^2$, M. Faux$^3$\\[1.5mm]
        B. Kleijn$^2$, R. Siebelink$^{1,4}$ and P. Termonia$^5$}\\
\vspace{7mm}
{\small
$^1$ Instituut voor Theoretische Fysica - Katholieke Universiteit Leuven\\
     Celestijnenlaan 200D B-3001 Leuven, Belgium\\[6pt]
$^2$ Institute for Theoretical Physics - Utrecht University\\
     Princetonplein 5, 3508 TA Utrecht, The Netherlands\\[6pt]
$^3$ Humboldt-Universit\"at zu Berlin, Institut f\"ur Physik\\
     D-10115 Berlin, Germany \\[6pt]
$^4$ Institute for Theoretical Physics, State University of New York at
     Stony Brook,\\ Stony Brook, NY 11790-3840, USA\\[6pt]
$^5$ Dipartimento di Fisica Teorica, Universit\'a di Torino\\
     via P. Giuria 1, I-10125 Torino\\
     Istituto Nazionale di Fisica Nucleare (INFN) -- Sezione di
     Torino, Italy  }
\end{center}
\vfill
\begin{center} {\bf Abstract} \\[3mm]
{\small
We review the coupling of $N=2$
supergravity to vector-tensor multiplets,\\ based on the method of
superconformal
multiplet calculus.}
\end{center}
\vfill
\flushleft{January 1998}
\end{titlepage}
%%%%%%%%%%%%%%%%%%%%%%%%%%%%%%%%%%%%%%%%%%%%%%%%%%%%%%%%%%%%%%%%%%%%%%%%%%%%%%
\def\titleline{Vector-Tensor Multiplets
%%%%%%%%%%%%%%%%%%%%%%%%%%%%%%%%%%%%%%%%%%%%%%
%                                            %
% Insert now the text of your title.         %
% Make a linebreak in the title with         %
%                                            %
%            \newtitleline                   %
%                                            %
%%%%%%%%%%%%%%%%%%%%%%%%%%%%%%%%%%%%%%%%%%%%%%
%%%%%%%%%%%%%%%%%%%%%%%%%%%%%%%%%%%%%%%%%%%%%%
}
\def\authors{P. Claus\1ad, B. de Wit\2ad, M. Faux\3ad, \newtitleline B.
Kleijn\2ad, R. Siebelink\1ad{}$^,$\4ad and P. Termonia\5ad
%%%%%%%%%%%%%%%%%%%%%%%%%%%%%%%%%%%%%%%%%%%%%%
%                                            %
%  Insert now the name (names) of the author %
%  (authors).                                %
%  In the case of several authors with       %
%  different addresses use e.g.              %
%                                            %
%             \1ad  , \2ad  etc.             %
%                                            %
%  to indicate that a given author has the   %
%  address number 1 , 2 , etc.               %
%                                            %
%%%%%%%%%%%%%%%%%%%%%%%%%%%%%%%%%%%%%%%%%%%%%%
%%%%%%%%%%%%%%%%%%%%%%%%%%%%%%%%%%%%%%%%%%%%%
}
\def\addresses{
%%%%%%%%%%%%%%%%%%%%%%%%%%%%%%%%%%%%%%%%%%%%%%
%                                            %
% List now the address. In the case of       %
% several addresses list them by numbers     %
% using e.g.                                 %
%                                            %
%             \1ad , \2ad   etc.             %
%                                            %
% to numerate address 1 , 2 , etc.           %
%                                            %
%%%%%%%%%%%%%%%%%%%%%%%%%%%%%%%%%%%%%%%%%%%%%%
\1ad Instituut voor theoretische fysica -- Katholieke Universiteit Leuven,\\
Celestijnenlaan 200D, B-3001 Leuven, Belgium\\
\2ad Institute for Theoretical Physics -- Utrecht University,\\
Princetonplein 5, 3508 TA Utrecht, The Netherlands\\
\3ad Humboldt-Universit\"at zu Berlin, Institut f\"ur Physik, D-10115
Berlin, Germany\\
\4ad Institute for Theoretical Physics, State University of New York at
Stony Brook,\\ Stony Brook, NY 11790-3840, USA\\
\5ad Dipartimento di Fisica Teorica, Universit\'a di Torino, via P. Giuria
1, I-10125 Torino\\
Istituto Nazionale di Fisica Nucleare (INFN) -- Sezione di Torino, Italy
%%%%%%%%%%%%%%%%%%%%%%%%%%%%%%%%%%%%%%%%%%%%%%%
}
\def\abstracttext{
%%%%%%%%%%%%%%%%%%%%%%%%%%%%%%%%%%%%%%%%%%%%%%%
%                                             %
% Insert now the text of your abstract.       %
%                                             %
%%%%%%%%%%%%%%%%%%%%%%%%%%%%%%%%%%%%%%%%%%%%%%%
We review the coupling of $N=2$
supergravity to vector-tensor multiplets, based on the method of
superconformal multiplet calculus.
%%%%%%%%%%%%%%%%%%%%%%%%%%%%%%%%%%%%%%%%%%%%%%%%%%%%%%%%%%%%%%%
}
\makefront
%%%%%%%%%%%%%%%%%%%%%%%%%%%%%%%%%%%%%%%%%%%%%%%%
%                                              %
%  Insert now the remaining parts of           %
%  your article.                               %
%                                              %
%%%%%%%%%%%%%%%%%%%%%%%%%%%%%%%%%%%%%%%%%%%%%%%%
\section{Introduction}\vskip-2mm
We review the couplings of vector-tensor multiplets to
supergravity and to generic configurations of vector multiplets 
\cite{vt1,vt2,vtsugra}, using the method
of superconformal multiplet calculus.  
The full supergravity theory
involving vector-tensor multiplets can be found in \cite{vtsugra} and an
extensive list of references is given there.  Here we explain the essential
features of the construction which could possibly have been obscured by the
many details given in \cite{vt1,vt2,vtsugra}.
Couplings of $N=2$ supergravity in four-dimensional spacetime
to a number of matter multiplets have already been studied extensively in the
past. The most well-known couplings involve vector
multiplets and hypermultiplets. Theories with $N=2$ supersymmetry
have provided a laboratory for exploring and testing many 
phenomena in perturbative and non-perturbative
field theory and in string theory (for some reviews, see e.g. \cite{SW}).

Besides the vector multiplet, the hypermultiplet and the vector-tensor
multiplet there exist two other multiplets that describe $4+4$ bosonic and
fermionic physical degrees of freedom.  They are the tensor multiplet and
the double-tensor multiplet.  All these multiplets can be described in
terms of 8+8 off-shell degrees of freedom. 
The fields of the vector-tensor multiplet are
a scalar $\phi$, a doublet of
Majorana spinors $\lambda_i$, a vector gauge field $V_\mu$, an
antisymmetric tensor gauge field $B_{\mu\nu}$ and an auxiliary scalar
$\phi^{(z)}$.  These fields are subject to an off-shell central charge,
which will be discussed later.
The interactions of the tensor gauge field
necessarily involve  Chern-Simons couplings to
vector gauge fields. Based on these Chern-Simons couplings
vector-tensor multiplets can be classified into two inequivalent classes.
When the tensor gauge field couples to its own vector $V_\mu$, the
transformation rules are non-linear in the components of the multiplet and
the action contains a self-coupling.  The other class is characterized by 
the fact that the tensor does not couple to its own vector and the 
transformation rules become linear in the vector-tensor fields.

Recently there has been quite some interest in setting up a suitable
superspace formulation for the vector-tensor multiplet.  In view of the complexity of
our results \cite{vtsugra}, a superspace description would be very welcome,
for instance to analyze possible non-renormalization properties of this
multiplet.  As a first step in this direction, the description of the
linearized vector-tensor multiplet in $N=2$ central-charge superspace 
was given in
terms of a constrained super 1-form \cite{HOW} and in terms of a
constrained super 2-form in \cite{GHH,BHO}.  In the last two papers the
linearized version of the multiplet with Chern-Simons couplings was
constructed.  However, the central charge gives rise to constraints on the
superfields, which have a rather complicated structure.

As is well known for the
massless hypermultiplet, one can circumvent an off-shell central
charge by adopting a description in terms of an infinite
number of fields.  A natural setting for this construction is provided by
harmonic superspace \cite{harmonic}, where hypermultiplets
are described by unconstrained analytic superfields.  The hope was
that a similar construction of the vector-tensor multiplet in harmonic
superspace would be possible.  However, it turned out that  
a central charge could not be avoided,
although the superfield constraints can be formulated more concisely in
harmonic superspace.  The harmonic superspace
version of the linearized multiplet was given in \cite{DK} and the
nonlinear version exhibiting a self-coupling was derived in \cite{DK2,IS}.
Gauging of the central charge in superspace has been considered in
\cite{DT}.  To date no full supergravity coupling in superspace has been
constructed.
%%%%%%%%%%%%%%%%%%%%%%%%%%%%%%%%%%%%%%%%%%%%%%%%%%%%%%%%%%%%%%%%%%%%%%%%%%%%%
\section{Superconformal multiplet calculus and the construction of N=2
supergravity-matter couplings}\vskip-2mm
In \cite{vtsugra} the method of superconformal multiplet calculus has been
reviewed, based on \cite{DWVHVP}. 
It allows a
systematic construction of off-shell supersymmetric Lagrangians in
terms of relatively small supermultiplets. These multiplets are
controlled by a large set of gauge transformations which contains the
superconformal group. Here we
demonstrate the method by showing how the field content of minimal
$N=2$ Poincar\'e supergravity is described in a gauge equivalent way
in terms of 2 superconformal multiplets.
One is the Weyl multiplet, which contains the gauge fields of the
superconformal algebra, the other an (abelian) vector multiplet.

The superconformal group in four dimensions is $SU(2,2|N)$ \cite{suconf}.
For $N=2$ supersymmetry the relevant algebra contains general-coordinate ($GCT$), local
Lorentz ($M$), scale ($D$), special conformal ($K$), chiral $U(1)$ and
$SU(2)$ and two types of supersymmetry transformations, called
$Q$-supersymmetry and $S$-supersymmetry
transformations.  The Weyl multiplet contains the corresponding gauge
fields and a number of auxiliary fields. These fields are
\arraycolsep=1mm
\begin{equation}
\begin{array}{rcccccccccccl}
(&e_\mu^a,&\underline{\omega_\mu^{ab}},&b_\mu,&\underline{f_\mu^a},&A_\mu,&
{\cal V}_{\mu i}^{\ \ j};&\psi_{\mu i},&\underline{\phi_{\mu
i}};&T^{ij}_{ab},&\chi_i,&D&)\\[4pt] &GCT&M_{ab}&D&K_a&U(1)&SU(2)^i_{\ \
j}&Q_i&S_i&&&&\\ &5&-&0&-&3&9&16&-&6&8&1&
\end{array}\,.\label{Weyl}
\end{equation}
The
fields that are underlined depend on
the other fields through conventional constraints
\cite{DWVHVP}.
These constraints involve the matter fields $T^{ij}_{ab}\, , \,\chi_i$ and $D$, such
that the multiplet has (24+24) off-shell degrees of freedom, as
specified in the
bottom line of (\ref{Weyl}).  
The fields given in (\ref{Weyl}) transform irreducibly under the
superconformal transformations.
The algebra associated with these transformations has 
field-dependent structure constants. The vierbein and the gravitini 
of the $N=2$ Poincar\'e
supergravity multiplet are contained in the Weyl
multiplet, 
but there is no candidate for the graviphoton.  The graviphoton 
in the gauge-equivalent description
resides in a superconformal \lq compensating\rq\,   vector multiplet
\begin{equation}
(X^0,\Omega^0_i,Y_{ij}^0,W_\mu^0)\,,\label{vector}
\end{equation}
which contains (8+8) off-shell degrees of freedom.  Because these two
multiplets contain the Poincar\'e supergravity fields, the resulting
field configuration is called the \lq minimal field representation\rq.  For a consistent 
gauge-equivalent formulation of the action, a second compensating multiplet
is required, e.g.  a hypermultiplet
\begin{equation}
(A_i^{\ \alpha},\zeta^\alpha,A_i^{\alpha(z)})\,,\label{hyper}
\end{equation}
which contains (8+8) off-shell degrees of freedom; here  $A_i^{\alpha(z)}$ is
an auxiliary field related to $A_i^{\ \alpha}$ by the action of
the central charge.  The other central-charge transformed fields
($\lambda_i^{(z)}, A_i^{\ \alpha(zz)}, ...$) all
depend algebraically on the fields given in (\ref{hyper}).  
We will comment on the
central-charge transformation in due course.

To construct a theory invariant under the superconformal group, we have to
impose scale, chiral $U(1)$ and $SU(2)$ invariance.  Therefore fields of
a superconformal multiplet acquire definite scaling and chiral $U(1)$
weights $(w,c)$,
\begin{equation}
\phi \rightarrow \exp[w\Lambda_D + ic \Lambda_{U(1)}]\,\phi\,.
\end{equation}
In the gauge-equivalent form, some of the multiplet components may act
as compensating fields. This has a profound effect on the
super-Poincar\'e theory. First of all, some of the superconformal 
gauge symmetries are no longer present, in particular the dilatations, 
the chiral $SU(2)$ and $U(1)$ transformations and $S$-supersymmetry. Secondly,
the Poincar\'e supersymmetry transformations take a more complicated
form and consist of a field-dependent linear combination of the original 
$Q$- and $S$-supersymmetry variations.
 
In the construction of theories with {\it rigid}
supersymmetry it is not imperative to impose invariance under scale
and chiral transformations. Nevertheless, this is what we did in our
initial construction of the coupling of vector-tensor to vector
multiplets.
This approach is tailor-made for constructing the coupling to
supergravity. However in the context of rigid supersymmetry, imposing
scale and chiral invariances does not amount to an extra restriction,
as it is always possible to freeze
the vector multiplet to a constant, i.e.
$(X^0,\Omega^0_i,Y_{ij}^0,W_\mu^0)\rightarrow (1,0,0,0)$. 

When based on a finite number of fields,
the off-shell representation of some multiplets
requires the presence of off-shell central charges.  
This is for instance the case for the
hypermultiplet used in (\ref{hyper}) and also for the vector-tensor
multiplet.  The central charge acts as a translation on the fields
generating an infinite sequence of central-charge transformed fields,
\begin{equation}
\phi\rightarrow \phi^{(z)}\rightarrow \phi^{(zz)}
\rightarrow \cdots\,.
\end{equation}
This central-charge transformation
appears in the commutator of two (rigid) supersymmetry transformations,
\begin{equation}
{}[\,\d_Q(\e_1),\d_Q(\e_2)\,]= \cdots + \delta_Z(z+\bar z)\,, \label{commu}
\end{equation}
with $z= 4 \ve^{ij} \bar \e_{2i} \e_{1j}$.
Imposing this algebra on the fields of the multiplet requires
constraints 
on the images of the fields under the central-charge 
transformation.  When we consider local
supersymmetry this central charge has to be realized in a local way as it
appears in the commutator of two local supersymmetry transformations.  So we have
to provide a gauge field for this symmetry, which is found in the
compensating vector multiplet (\ref{vector}) of the minimal field
representation.  However, on this multiplet the supersymmetry algebra closes
into a field-dependent gauge transformation (which is identified with the central-charge
transformation).  For $W_\mu^0$ we have
\begin{equation}
{}[\,\d_Q(\e_1),\d_Q(\e_2)\,] W_\mu^0 = \cdots + \partial_\mu (z+\bar z)\,,
\end{equation}
where now $z= 4\ve^{ij} \bar \e_{2i} \e_{1j} X^0$.
The key observation here is that realizing the central charge in a
local fashion, as is required for local supersymmetry, gives rise to a
field-dependent central-charge transformation in the commutator of two
supersymmetries. With respect to the super-Poincar\'e transformations the
Weyl multiplet and the compensating vector-multiplet define an off-shell
irreducible multiplet, known as the minimal field representation and
consisting of 32+32 degrees of freedom.

The Poincar\'e supergravity action can be conveniently derived by adding the
compensating hypermultiplet and imposing
gauge-conditions on the components of the compensating fields,
\begin{equation}
b_\mu=0;\quad X^0=\kappa^{-1};\quad \Omega_i^0=0;\quad
A_i^{\ \alpha}=\delta_i^\alpha ;\quad \zeta^\alpha=0\,,\label{gauge}
\end{equation}
where $\kappa$ is the gravitational coupling constant.
The fields $T^{ij}_{ab}, \chi_i, D, A_\mu, {\cal V}_{\mu i}^{\ \ j},
Y_{ij}^0$ and $A_i^{\alpha(z)}$ are auxiliary and can be eliminated as 
independent fields.  In this way, we end up with the on-shell
$N=2$ Poincar\'e supergravity multiplet with the fields
$
(e_\mu^a,\psi_{\mu i},W_\mu^0)\,,
$
associated with the graviton, gravitini and graviphoton states.
The conditions (\ref{gauge}) freeze most of the components of the
compensating multiplets to constants, and break the scale, chiral $U(1)$, and
$SU(2)$ transformations. 
Having discussed some of the properties of the
superconformal approach for constructing supergravity-matter
couplings, let us summarize the steps of the construction of
supergravity-matter couplings in the superconformal approach.  Firstly we demand invariance
under rigid scale and chiral $U(1)$ transformations and we gauge the
central charge in rigid supersymmetry, i.e.~we introduce the compensating
vector multiplet as the gauge-vector multiplet of the central-charge
transformation.  
Secondly, we impose the superconformal algebra on all the fields. This
is usually done by iteration. Thirdly, we construct Lagrangians
invariant under all superconformal transformations, by 
using certain density formulas.  After this step one 
exhibits the nature of the gauge-equivalent formulation in which some
of the superconformal symmetries are absent. This is most conveniently done by
imposing gauge choices on the compensating fields.
%%%%%%%%%%%%%%%%%%%%%%%%%%%%%%%%%%%%%%%%%%%%%%%%%%%%%%%%%%%%%%%%%%%%%%%%%%%%%%
\section{Vector-tensor multiplets coupled to N=2 supergravity}\vskip-2mm
In this section we apply the method outlined above to the
supergravity coupling of vector-tensor multiplets.  The linearized
transformation rules for this multiplet under rigid supersymmetry were
given in \cite{sohnius,DWKLL} and it was shown that the (rigid) algebra
closes into an off-shell central-charge transformation.  A finite number of
degrees of freedom describing the multiplet consists of a scalar $\phi$, a doublet
of Majorana spinors $\lambda_i$, a vector gauge field $V_\mu$, with field
strength $F_{\mu\nu}$, an antisymmetric tensor gauge field $B_{\mu\nu}$,
with field strength $H_{\mu\nu\rho}$ and an auxiliary scalar $\phi^{(z)}$.
With the linearized transformation rules given in \cite{DWKLL}, the commutator of two
(rigid) supersymmetries on $B_{\mu\nu}$ yields
\brr
{}[\,\d_Q(\e_1),\d_Q(\e_2)\,] B_{\mu\nu}&=& (2 \bar \e_{2}^i \gamma^\rho
\e_{1i} + {\rm h.c.})(\partial_\rho B_{\mu\nu}
+ 2  \partial_{[\mu} B_{\nu]\rho} ) \nonumber\\
&&
- 2 \partial_{[\mu}(\bar \e^i_{2} \gamma_{\nu]} \e_{1i} \phi +
{\rm h.c.}) -\ft i4 (z + \bar z)\ve_{\mu\nu\rho\sigma}F^{\rho\sigma}
+ \ft i2(z-\bar z) F_{\mu\nu}\,. \label{QQB}
\err
The first, second and third term on the right-hand side of this
equation represent an infinitesimal
translation, a tensor gauge transformation and a central-charge
transformation, respectively.  The complex parameter $z$ is equal to
$4\ve^{ij} \bar \e_{2i}\e_{1j}$, which  is  constant at
this stage. 
At the linearized level the last term can be interpreted in two ways.
For the first interpretation, one moves  $z-\bar z$ under the
derivative. The transformation on $B_{\mu\nu}$ then takes the form of a
tensor-gauge transformation
\begin{equation}
\delta_\Lambda B_{\mu\nu} = 2 \partial_{[\mu} (\ft i2 (z-\bar z) V_{\nu]})\,.
\end{equation}
However, when requiring scale and chiral invariance of the
transformation rules of the vector-tensor multiplet, the parameter $z$
acquires a field-dependent modification 
$z=4\ve^{ij} \bar \e_{2i}\e_{1j} X^0$. Also for a locally
supersymmetric version of the vector-tensor multiplet, $z$ becomes
spacetime dependent through the supersymmetry parameters $\e_{i}$. 
In those cases of a non-constant $z$, the last term in (\ref{QQB})     
can no longer be interpreted as a tensor-gauge transformation and can only be
identified as a new gauge transformation associated with a Chern-Simons coupling between
$B_{\mu\nu}$ and $V_\mu$,
\begin{equation}
\delta_{\theta^1} B_{\mu\nu} =
i (z-\bar z) \partial_{[\mu} V_{\nu]}\,.\label{CSD}
\end{equation}
The result is that gauging the central charge in the
supersymmetric theory leads necessarily to a Chern-Simons coupling for the
tensor field.  Demanding that the multiplet has balanced scaling and chiral
$U(1)$ weights forces the fields of the vector multiplet to act
as compensators.  The minimal configuration that only involves the
vector-tensor multiplet and the central-charge vector multiplet was
constructed in \cite{vt1}.  The direct Chern-Simons coupling (\ref{CSD}) of
the vector and the tensor of the vector-tensor multiplet leads to
unavoidable non-linearities of the action and the transformation rules.
Freezing the vector multiplet to a constant and after suitable rescalings
of the fields, its rigidly supersymmetric  action takes the form
\brr\label{vt-selfinteraction}
{\cal L}&\propto& \ft12 \phi (\der_\m\phi)^2
+ \ft14 \phi (\der_\m V_\n-\der_\n V_\m)^2 +\ft34 \phi^{-1}
\Big(\der_{[\m}B_{\n\rho]}- V_{[\m}\der_\n V_{\rho]}\Big)^2 \nonumber \\ &&
+\ft12 \phi\, \bar \l^i {\stackrel{\leftrightarrow}{\dslash}} \l_i -2\phi\,
(\phi^{({\rm z})})^2- \ft14 i\Big(\ve^{ij} \bar \l_i\s^{\m\n} \l_j - {\rm
h.c.}\Big) \, (\der_\m V_\n - \der_\n V_\m) \nonumber\\ && -\ft1{24}
\phi^{-1} \bar \l_i \g^\m \g^\n\g^\rho \l^i \Big(\der_{[\m} B_{\n\rho]} -
V_{[\m} \der_\n V_{\rho]} \Big) + \ft3{16} \phi^{-1}
\Big(\bar\l^i\g_\m\l_i\Big)^2 \nonumber\\ && + \ft1{32}
\phi^{-1}\Big((\ve^{ij} \bar\l_i\s_{\m\n}\l_j)^2 + {\rm h.c.} \Big) \,.
\err

Having established this non-linear version, a more systematic analysis of
the multiplet was made by introducing a general (up to field redefinitions)
Chern-Simons coupling of the tensor to a generic background configuration
of vector multiplets in \cite{vt2}.  The transformation of the total
field configuration under the gauge symmetry associated with $V_\mu$
($\delta_{\theta}(\theta^1)$), the tensor gauge symmetry $\delta_\Lambda
(\Lambda_\mu)$, the central-charge transformation $\delta_Z(z+\bar z)$ and
the background gauge symmetries $\delta_{\theta}(\theta^A)$ is given by
\brr
\d W_\mu^0&=& \partial_\mu(z + \bar z)\,,\qquad
\d W_\mu^A = \partial_\mu \theta^A\,,\nonumber\\
\d \phi &=& (z+\bar z) \phi^{(z)}\,,\qquad
\d \phi^{(z)}=(z+\bar z)\phi^{(zz)}\,,\quad
\d \lambda_i = (z + \bar z) \lambda_i^{(z)}\,,\nonumber\\
\d V_\mu &=& \partial_\mu \theta^1 + (z+\bar z) V_\mu^{(z)}\,,\nonumber\\
\d B_{\mu\nu} &=& 2 \partial_{[\mu} \Lambda_{\nu]} + \eta_{11} \theta^1
\partial_{[\mu} V_{\nu]} + \eta_{1A} \theta^1 \partial_{[\mu} W_{\nu]}^A +
\eta_{(AB)} \theta^A \partial_{[\mu}W_{\nu]}^B \nonumber\\
&&+ (z + \bar z) [- \eta_{11} V_{[\mu} V_{\nu]}^{(z)} + \hat
B_{\mu\nu}^{(z)}]\,.
\err
where $\eta_{11}$, $\eta_{1A}$ and $\eta_{AB}$ are constants.

The algebra constrains $\lambda_i^{(z)}$, $\phi^{(zz)}$ in
terms of the other fields and
\brr V_\mu^{(z)}&=& - \frac{1}{G(\phi, X^I, \bar X^I)} H_\mu
+ E(\phi,X^I,\bar X^I)+\dots\,,\nonumber\\[5pt]
\hat B_{\mu\nu}^{(z)} &=& A(\phi,X^I, \bar X^I) F_{\mu\nu}
+ B(\phi,X^I, \bar X^I)\ft12 \ve_{\mu\nu\rho\sigma} F^{\rho\sigma} \nonumber\\
&&+ C_I(\phi,X^I, \bar X^I) \F^I_{\mu\nu}
+ D_I(\phi,X^I, \bar X^I) \ft12 \ve_{\mu\nu\rho\sigma} \F^{I\rho\sigma}
+ \dots \,,
\err
where the functions $A,B,C_I,D_I, G$ and $E$ are known and the dots
indicate fermion terms.  {}From $\hat B_{\mu\nu}^{(z)}$ we can see that the
central charge induces field-dependent Chern-Simons couplings to the
background vector multiplets.  This indicates that the gauge field
associated with the central charge will exhibit rather peculiar
(nonpolynomial) couplings (see also \cite{DT,DWVHVP}).  Closure of the
supersymmetry algebra on the fields yields the following interesting
result: it is not possible to put both $\eta_{11}$ and $\eta_{1A}$ equal to
zero at the same time.  This results in two classes of inequivalent
vector-tensor multiplets.  One version with $\eta_{11}\ne 0$ and
$\eta_{1A}$ gives non-linear transformation rules for the vector-tensor
multiplet and a self-interacting theory, as given in
(\ref{vt-selfinteraction}) when the background is frozen.  The other
version $\eta_{11}=0$ and $\eta_{1A}\ne0$ needs besides the central-charge
vector multiplet one additional vector multplet and results in a linear
version of the vector-tensor multiplets, with no self-interaction.  The
coupling of both versions to $N=2$ supergravity has been performed in
\cite{vtsugra} and no new features appear.  There are still two
inequivalent versions which are encoded in specific Chern-Simons terms.
%%%%%%%%%%%%%%%%%%%%%%%%%%%%%%%%%%%%%%%%%%%%%%%%%%%%%%%%%%%%%%%%%%%%%%%%%%%
\section{Conclusions}\vskip-2mm
We have reviewed some essential features of the coupling of vector-tensor
multiplets to $N=2$ supergravity using the superconformal multiplet
calculus.  This leads us to two inequivalent versions of this multiplet
with Chern-Simons couplings of the tensor to its own vector
and to a generic configuration of (background) vector multiplets
\cite{vt2}.

In four spacetime dimensions one can dualize an
 antisymmetric tensor to a
scalar $a$. Together with $\phi$ it combines to the scalar of a vector
multiplet $X^1=X^0(a + i\phi)$.  Vector-multiplet couplings in $N=2$
supergravity can be
characterized by a prepotential $F$.  This prepotential is 
intimately related to the Chern-Simons couplings.  Summarizing we have the
following two versions:\\
{\it the non-linear vector-tensor multiplet}
\begin{equation}
\d B_{\mu\nu} = \eta_{11} \theta^1 \partial_{[\mu}
V_{\nu]} + \eta_{AB} \theta^A \partial_{[\mu} W_{\nu]}^B \rightarrow F=
\eta_{11}\frac{(X^1)^3}{X^0} + \eta_{AB} \frac {X^1}{X^0} X^A X^B
\end{equation}
and {\it the linear vector-tensor multiplet}
\begin{equation}
\d B_{\mu\nu} = \eta_{1A} \theta^1 \partial_{[\mu}
W_{\nu]}^A + \eta_{AB} \theta^A \partial_{[\mu} W_{\nu]}^B \rightarrow F=
\eta_{1A}\frac{(X^1)^2}{X^0} X^A + \eta_{AB} \frac {X^1}{X^0} X^A X^B \,.
\end{equation}
Possible applications of vector-tensor multiplets in string
theory have been commented on in \cite{vtsugra} and we refer the interested
reader to references given there.
\vskip5mm
\noindent
{\bf Acknowledgements:}\ \ This work is
supported by the European Commission TMR programme ERBFMRX-CT96-0045, in
which P.C. is associated to Leuven, B.d.W en B.K. to Utrecht, M.F. to Berlin
and P.T. to Torino.

\end{document}